# Text-aided Group Decision-making Process Observation Method (x-GDP): A novel methodology for observing the joint decision-making process of travel choices

Giancarlos Parady  Yuki Oyama  Makoto Chikaraishi
The University of Tokyo  Shibaura Institute of Technology  Hiroshima University

Keywords: *Joint decision-making, social networks, leisure travel behavior, group behavior, survey methods*

## Abstract

Joint travel decisions, particularly related to social activities remain poorly explained in traditional behavioral models. A key reason for this is the lack of empirical data, and the difficulties associated with collecting such data in the first place. To address this problem, we introduce Text-aided Group Decision-making Process Observation Method (x-GDP), a novel survey methodology to collect data on joint leisure activities, from all members of a given clique. Through this method we are able to observe not only the outcome (i.e., the joint activity location chosen) but also the decision-making process itself, including the alternatives that compose the choice set, individual and clique characteristics that might affect the choice process, as well as the discussion behind the choice via texts. Observing such a process will allow researchers to gain a deeper understanding of the joint decision-making process, including how alternatives are weighted, how members interact with each other, and finally how joint choices are made. In this paper we introduce the results of a x-GDP survey implementation focusing on joint eating-out activities in the Greater Tokyo Area, giving a detailed overview of the survey components, execution logistics and initial insights on the data. This is to the best of our knowledge the first attempt to observe group joint travel decisions in real time through a zoom-moderated experiment.

## 1. Introduction

As social animals, many of our behavioral decisions, including travel-related ones, are made in coordination with members of the social networks we are embedded in. However, joint decision-making processes, particularly related to social activities remain poorly explained in traditional behavioral models. A key reason for this is the lack of empirical data, and the difficulties associated with collecting such data in the first place. While some studies have indeed focused on modeling joint activities, these studies rely on agent-based simulations (Arentze and Timmermans, 2008; Ronald, Arentze and Timmermans, 2012) and still require empirical data for parameter estimation and model validation.

In recent years, several egocentric network data-collection efforts have been conducted to get a better understanding of ego-centric social networks characteristics and social interactions (Parady *et al.*, 2021). Such efforts include similar surveys in Canada (Carrasco and Miller, 2006), Switzerland (Frei and Axhausen, 2008; Kowald and Axhausen, 2012; Guidon *et al.*, 2018), The Netherlands (van den Berg, Arentze and Timmermans, 2012), Chile (Carrasco and Cid-Aguayo, 2012) and Japan

(Parady, Takami and Harata, 2020) and the U.K. (Calastri, Crastes dit Sourd and Hess, 2020). A key limitation of these efforts is that since data is collected using an ego-centric approach, the data that can be collected on alters (other group members) is limited to what ego can observe and recall. This limitation is particularly critical for modeling travel behavior as spatio-temporal constraints are a key set of constraints defining travel behavior (Hagerstrand, 1970) . Han *et al.* (2023) has shown in the context of group eating-out destination choices that considering the average travel times of all participating members of a clique increases the predictive ability of the model by up to 49% against a model considering only ego's travel times, a considerable increase in performance, especially given there has been very little progress in improving the predictive ability of destination choice models in the past decades.

Against this background, in this article we introduce x-GDP (Text-aided Group Decision-making Process Observation Method), a novel survey method to collect data on joint activities and their underlying joint decision-making processes. We applied the method to joint leisure activities with a particular focus on destination choice, but this proposed method can be easily generalized to other dimensions of travel behavior. Through this method we are able to observe not only the outcome (i.e., the final location chosen) but also the decision-making process itself, including the alternatives that compose the choice set, individual and clique characteristics that might affect the choice process, as well as the discussion behind the choice via texts. Observing such a process will allow us to first understand the decision-making process qualitatively, including how alternatives are weighted, how members interact with each other, and finally how the choice is made.

## 2. Text-aided Group Decision-making Process Observation Method (x-GDP): an overview

The main objective of this method is to collect real-time data on the joint decision-making process of travel-related activities of a given clique, a group where all members know each other. In a nutshell, the general idea of x-GDP is to ask participant cliques to plan (and later actually execute) an actual activity or set of activities in the virtual presence of the researchers, using a chat-group interface. In this paper, we use as a case study eating-out activities, because eating-out is the most frequently executed joint-leisure activity (Stauffacher *et al.*, 2005). A key aspect of this method is that since participants have to actually conduct the activity decided in the group discussion (proof of execution is one of the conditions to receive the participation monetary incentive) there are incentives in place to guarantee a real discussion that takes into consideration the preferences and constraints of clique members.

Fig 1 illustrates the flow of an x-GDP survey. In particular, the crux of the method is Step 3, executed over a virtual meeting (via any market-available web-conference app). The widespread of tele-commuting and tele-conferencing resulting from the COVID-19 pandemic made such a virtual implementation much easier to implement as people are now very familiar with such meeting tools.

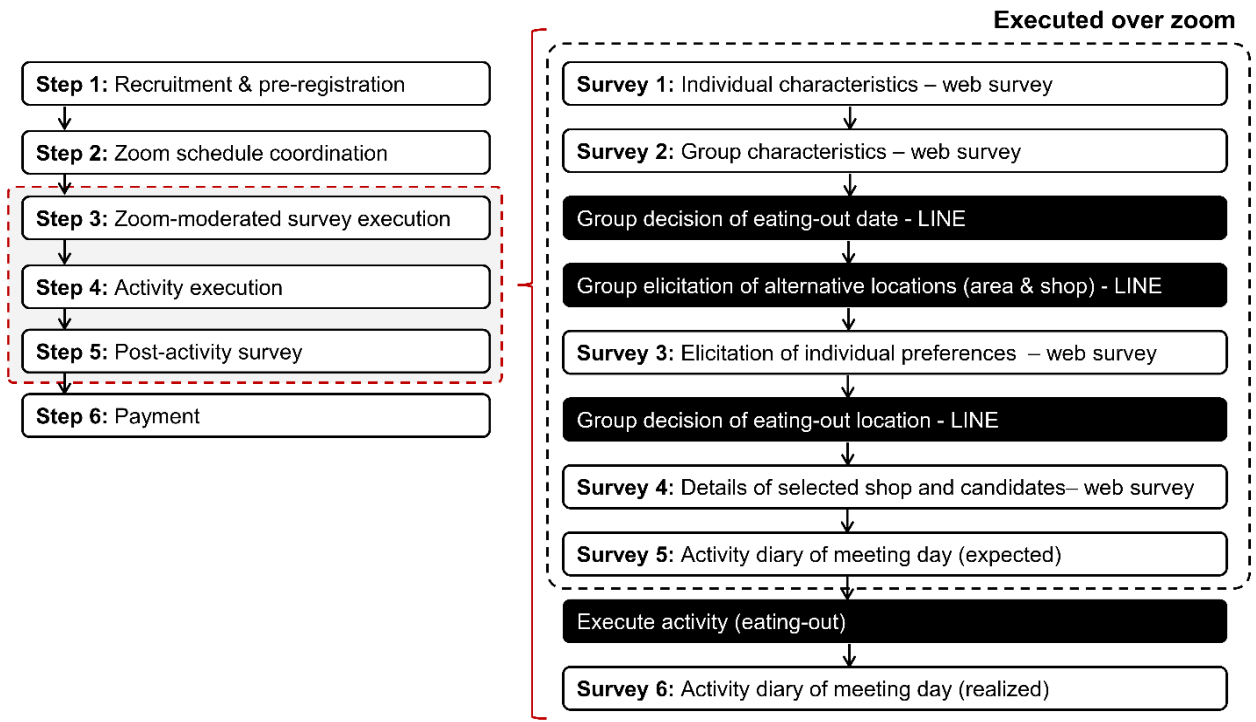

Fig 1. Flow of an x-GDP survey

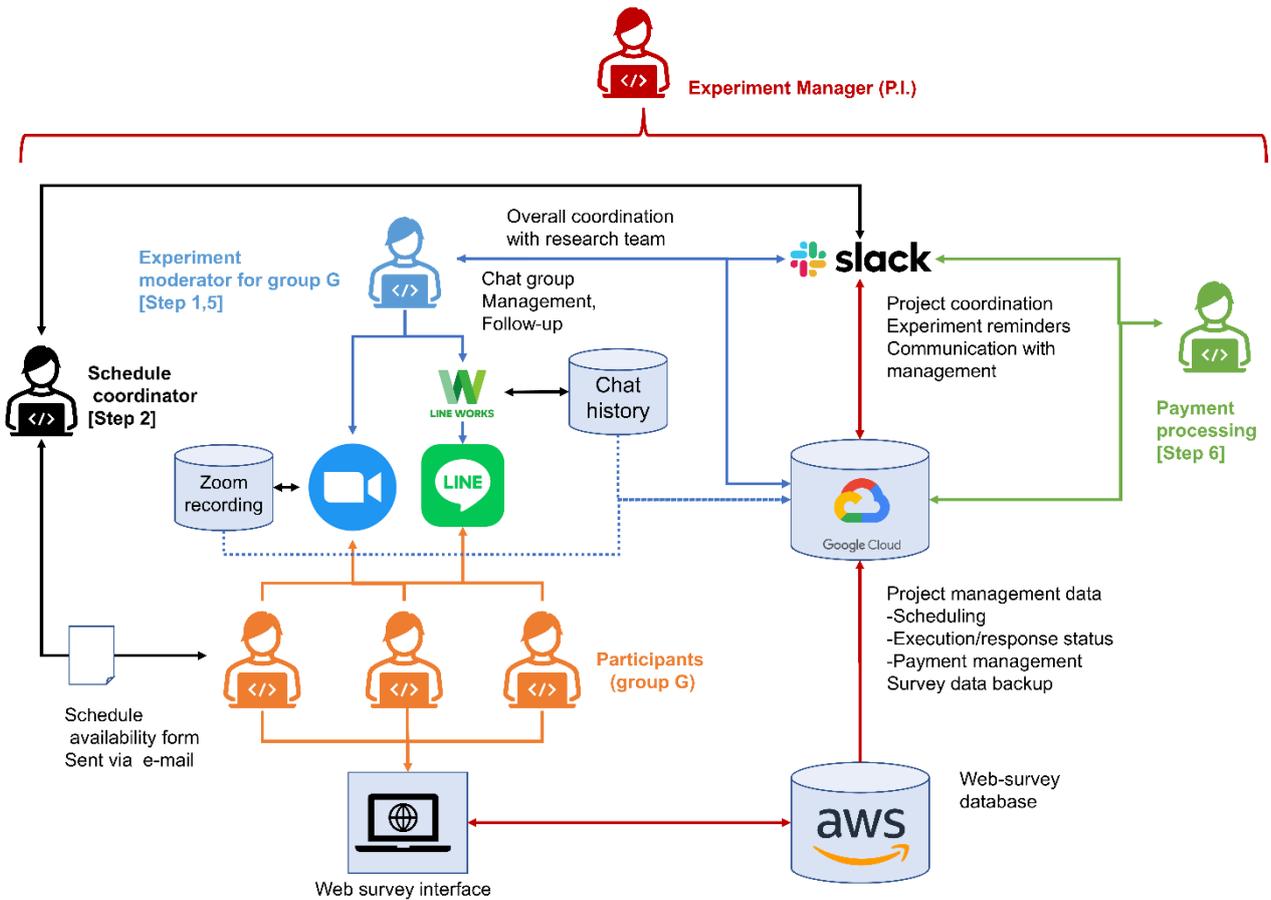

Fig 2. Simplified diagram of the logistics of the x-GDP experiment after recruitment.

## 2.1. Step 1: Recruitment and pre-registration

x-GDP requires the participation of existing cliques. As such, registration of all members is necessary for schedule coordination. Since no such sampling frames exist of cliques, there were some challenges to sample recruitment. In the experiment presented in this study, we targeted cliques composed of at least one University of Tokyo student to simplify the sampling process. This was also done to limit to some extent the spatial distribution of participants to cliques with similar daily life activity spaces (i.e. avoid sampling cliques with very sparse and spread out geographical distributions that would make destination choice modeling difficult). Provided this condition was met, no constraints were imposed on the eligibility of other members.

Recruitment was done via social media (the Urban Transportation Research Unit Twitter account). In spite of the nonprobability sampling method, it is important to point out that the student population of the University of Tokyo is not that large (27,233 students as of November 2022) and is rather homogeneous in terms of sociodemographics. As such, we believe we were able to capture a set of participants that resemble the general characteristics of the student body of the university. Given this fact, and the experimental nature of this study, the current sampling method was judged adequate. In total, data on 816 individuals, belonging to 217 cliques was collected Out of the 816 participants 76% were University of Tokyo students, 20% were students from other universities and 4% were non-students (See the appendix for a summary of individual sociodemographic characteristics of the sample).

## 2.2. Step 2: Virtual meeting schedule coordination

Once registration of all cliques was done, scheduling coordination was conducted via online forms. As shown in Fig 2, the Schedule Coordinator matched Experiment Moderators (the person in charge of moderating and guiding the experiment over Zoom) with cliques based on their availability. In the experiment presented in this study, schedules were allocated in 2-week phases and constrained by the availability of Moderators. If a particular clique remained unmatched during a particular phase, they were reassigned to subsequent phases by the Schedule Coordinator.

Once a clique was successfully matched, all members were informed of the date and time of the experiment, as well as the Zoom link. In addition, detailed experiment explanations (including conditions for payment of participation reward) and informed consent forms were also sent. Scheduling was a rather challenging task given the need to coordinate time for all clique members and the Experiment Moderators (in total 4 to 7 persons). Early-morning (before school or work) and late evening (after school or work) were popular time slots.

## 2.3. Step 3: Zoom-moderated survey execution

The zoom-moderated experiment was executed as scheduled in Step 2. This was the crux of the experiment. (see Fig 2 for the logistics of this step). Although explanation of the experiment and informed consent forms were sent beforehand, at the beginning of the experiment the Experiment Moderator explained verbally the details of the survey as well as the conditions for the payment of the participation reward.

Guided by the Experiment Moderator, participants were first asked to respond to Survey 1 and Survey 2 via an online survey platform developed in-house. The Experiment Moderators shared via Zoom chat the link to the questionnaires as well as login information. Survey 1 collected data on individual socio-demographic characteristics (see the appendix for a summary of individual level characteristics of the sample). Survey 2 collected data on clique characteristic and was answered at the clique-level. To do so, the Experiment Moderator asked one member to share his/her screen while answering the survey. Participants could freely speak during this survey.

After Survey 2 was completed, the Experiment Moderator invited all members to a LINE group chat (LINE is the most popular instant communication freeware app in Japan, and we correctly anticipated that all participants would already have an account by the time of the experiment). Participants joined the LINE group chat on their personal accounts; however, the Experiment Moderator joined via a Line Works account (a cloud-based business chat tool that can link to LINE). This was done for privacy and ethical reasons as well as data management reasons. By connecting via a Line Works interface, data was not saved in the Experiment Moderators' personal accounts. Instead, it was saved centrally in the Line Works database, which the Experiment Managers can control access to. This also protected the privacy of the Experiment Moderators who needed not share their personal account information.

In the LINE group chat, the Experiment Moderator asked the clique to first decide the date and time of the activity. In the study presented in this article, two scheduling constraints were imposed. First, for management reasons, the date of the activity must be within a maximum of two weeks from the day of the experiment (three weeks was allowed at the discretion of the Experiment Moderator if no consensus was reached). The second constraint was that the activity must be done from the evening on (from 17:00~). This was done to reduce the temporal variability of the activities and simplify the modeling process later on. Note that these constraints can be easily generalized depending on the activity or model of interest to the researcher.

Once the date and time for the activity were defined, participants were asked to elicit potential areas and shops to execute the activity. There was no upper bound on how many candidates could be elicited but participants were asked to propose at least one location per person. Before moving on to the discussion phase to choose the activity location, respondents were asked to respond to Survey 3, which asked them to rank the elicited candidate locations in order of their personal preference. This was done anonymously so that responses were not affected by the opinions of others.

After completing Survey 3, participants were asked to discuss and decide the location of the eating-out activity. No guidance was given regarding how to make this decision, so each clique was free to choose their own method. There was no time constraint imposed on the LINE group discussion. Once a decision was made, the LINE group discussion part of the experiment was completed. The average duration for the LINE discussion section including time decision, preference elicitation and location decision was 35 minutes (S.D. 16.42 mins). The moderator asked participants to respond to Survey 4 via a web-survey (at the clique level). This survey collected data on the chosen location as well other candidate locations considered. To avoid the issue of untraceable locations, participants were asked to use store hyperlinks from either Tabelog (a restaurant review site in Japan) or Google maps. Other links were allowed if and only if the store did not have either a Tabelog or a google place links. Out of the 1,188 unique shops elicited during the course of the experiment, we were able to identify 1,182 (99.5%) of the shops via their public links and collect additional data on these shops.

Finally, once Survey 4 was completed, each participant was asked to report their expected schedule for the day of the activity in the form of an activity diary (Survey 5). This was done via a visual and interactive interface that greatly reduces the response burden (See the appendix for a screenshot of the survey interface).

## 2.4. Step 4: Activity execution

On the morning of the day of the planned activity, participants were sent a reminder via LINE and were given explanations about proof-of-execution. More specifically, respondents were asked to:

1. Share their location during the activity via LINE
2. Take a picture in front of the shop along with a mobile phone showing date and time
3. Take a group picture inside the restaurant
4. Receipt

## 2.5. Step 5: Post-activity survey

Using the same interface as Survey 5, data was collected on the actual schedule executed on the day of the activity. To reduce the response burden, their responses for Survey 5 (i.e., the expected schedule) were presented first and respondents could edit these when changes in the scheduled had occurred.

## 2.6. Step 6: Payment

A monetary incentive of JPY 4000 (approx. USD 29.80 as of Feb. 20, 2023) for participants who responded to all surveys and provided proof-of-execution. Irrespective of reason, for participants who did not provide proof-of-execution (including non-participation) or did not complete Survey 6 after participation, the incentive was JPY 1080 (approx. US$8).

## 3. Data characteristics and preliminary findings

### 3.1. Clique characteristics

Table 1 summarizes the clique-level characteristics of the sample. Clique size was limited by design so it ranges from three to five persons. Relationship length was measured for all the possible dyads within the clique (1,191 dyads). For the majority of the dyads, the relationship length is longer than three years. As expected, eating out constitutes the most common joint activity of all cliques (44.7%) followed by hobbies (21.66%) and university club and circle activities (17.05%) such as sports, arts, etc. In terms of within-clique hierarchy, 72.4% of cliques had no hierarchy. Hierarchy was self-reported by cliques, but in the context of Japanese social networks, hierarchy is usually defined as age difference or grade differences. Two-level hierarchy cliques accounted for 18.4% of cliques. In such cliques there are only two degrees of hierarchy (for example, all members are either undergraduate 1st year or undergraduate 2nd year).

Contact frequency was measured as meeting or ICT-mediated frequency of at least three members of the clique (contact between dyads are not considered). 12% of cliques meet once to three times per week, but the majority meets roughly once a month or less often.

Table 1. Clique level characteristics of sample (n=217)

| Variable | n | Share (%) |
|---|---|---|
| Clique size | | |
| 3 persons | 97 | 45% |
| 4 persons | 75 | 35% |
| 5 persons | 45 | 21% |
| Relationship length (time knowing each other) | | |
| Less than 1 year | 182 | 15.30% |
| 1-3 years | 414 | 34.80% |
| 3-5 years | 270 | 22.70% |
| More than 5 years | 325 | 27.30% |
| Main joint activity of clique | | |
| Eating-out | 97 | 44.70% |
| Hobby | 47 | 21.66% |
| Club/circle activities | 37 | 17.05% |
| Education/research | 28 | 12.90% |
| Job-related | 3 | 1.38% |
| Others | 5 | 2.30% |
| Hierarchical structure | | |
| No hierarchy | 157 | 72.4% |
| Two-level hierarchy | 40 | 18.4% |
| Three-level hierarchy | 18 | 8.3% |
| Four-level hierarchy | 2 | 0.9% |
| Frequency of face-to-face communication | | |
| Everyday | 3 | 1.40% |
| 4-6 per week | 17 | 7.80% |
| 1-3 per week | 26 | 12.00% |
| 3-4 per month | 12 | 5.50% |
| 1-2 per month | 30 | 13.80% |
| 4-11 per year | 47 | 21.70% |
| 1-3 per year | 56 | 25.80% |
| Once a year or less | 26 | 12.00% |
| Frequency of communication through SNS | | |
| Everyday | 24 | 11.06% |
| 4-6 per week | 23 | 10.60% |
| 1-3 per week | 44 | 20.28% |
| 3-4 per month | 27 | 12.44% |
| 1-2 per month | 34 | 15.67% |
| 4-11 per year | 34 | 15.67% |
| 1-3 per year | 12 | 5.53% |
| Once a year or less | 4 | 1.84% |
| No contact at all | 15 | 3.92% |
| Eating-out frequency by clique | | |
| Everyday | 0 | 0.00% |
| 4-6 per week | 4 | 1.84% |
| 1-3 per week | 6 | 2.76% |
| 3-4 per month | 7 | 3.23% |
| 1-2 per month | 36 | 16.59% |
| 4-11 per year | 60 | 27.65% |
| 1-3 per year | 65 | 29.95% |
| Once a year or less | 39 | 17.97% |

### 3.2. Event characteristics

Fig 3 illustrates both the chosen restaurant location as well as other considered candidates. The first thing to point out is the agglomeration of locations around Tokyo sub-centers such as Shibuya, Shinjuku, Ikebukuro, Ueno, Tokyo and Ginza connected via the Yamanote loop line, in addition to areas around the University of Tokyo's Komaba and Hongo Campus. Historically, the Tokyo sub-centers have exhibited high degrees of agglomeration of commercial and other facilities due to their high levels of access both from the railway-connected suburbs as well as the city center. In addition, smaller agglomerations can be seen around the intersection of railway lines even though they are not central.

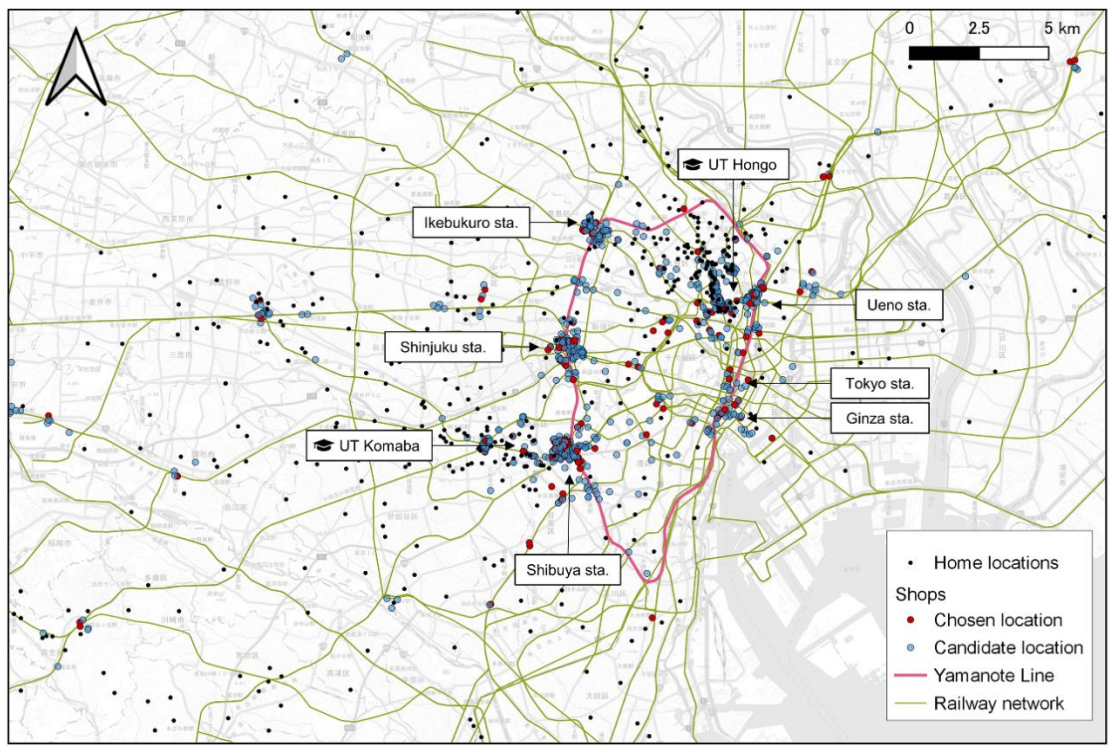

Fig 3. Location of chosen restaurants and alternatives considered during the experiment.

Table 2 summarizes the characteristics of the scheduled events. As mentioned above, the start time of the joint eating-out (hereinafter, the main activity) was constrained by design to after 17:00. 80.6% of all activities were conducted on weekdays.

Table 2. Characteristics of scheduled joint eating-out activity (n=217)

| Variable | n | Share (%) |
| --- | --- | --- |
| Scheduled activity start time | | |
|     17:00~17:59 | 13 | 6.0% |
|     18:00~18:59 | 54 | 24.9% |
|     19:00~19:59 | 111 | 51.2% |
|     20:00~20:59 | 21 | 9.6% |
|     21:00~21:59 | 15 | 6.9% |
|     22:00~ | 3 | 1.4% |
| Day of week | | |
|     Weekend | 42 | 19.4% |
|     Weekday | 175 | 80.6% |

Fig 4 shows the origin-destination (OD) distance distribution for the trips towards the main activity as reported by the clique on Survey 4, and the trips towards the individually top-ranked location as reported in Survey 3. We hypothesized that individuals would highly rank locations that are closer to them because the preference elicitation survey (Survey 3) was anonymous. However, contrary to our expectations, there were no large differences at the aggregate level between OD distances for the chosen restaurant by the clique and for the individually top-ranked restaurants. The median difference was 450 meters.

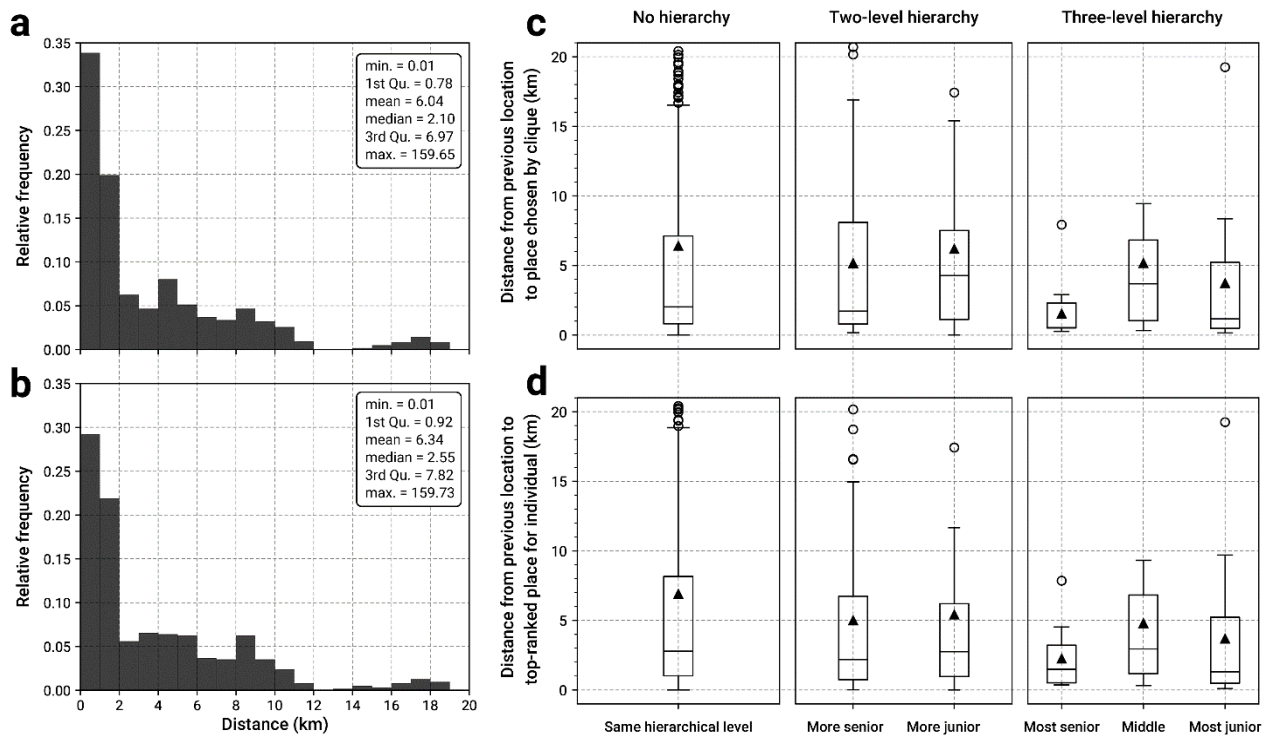

Fig 4. OD distance distribution of trips towards restaurant chosen by clique (a) and its respective box plot by hierarchy (c). OD distance distribution of trips towards the individually top-ranked location (b) and its respective box plot by hierarchy (d)

Regarding differences by hierarchy level, we hypothesized that more junior members of the clique would defer to the more senior members' preferences in terms of trip distance, however, although a weak association between hierarchy and trip distance was observed, it was not as strong as we expected. Note that four-level hierarchy was omitted from the figure, because there were only 2 cases out of 217 that fell in this category.

### 3.3. Decision-making process characteristics

Table 3 summarizes the reasons for choosing restaurant for the main activity. These data were categorized from free-answers in Survey 4, where the clique was asked to briefly summarize the main reasons behind their choice. As expected, restaurant quality and accessibility were the most frequently mentioned factors (78.8% and 57.1% of clique mentioning them, respectively). This is also consistent with the attitudinal responses collected in the individual survey (Survey 1) where respondents were asked to rate on a 7-point Likert scale (1 being not important at all, 7 being extremely important), the importance they place on different factors when eating out with a group (Fig 5). Group evaluation of shop and group transit access were rated six or seven by 71.2% and 76.7% of the individuals, respectively. In contrast, individual evaluation of shop and individual transit access were rated six or seven by 65.7% and 59.3% of the individuals. This might explain why no large difference were observed in Fig.4 between OD distances for the chosen restaurant by the clique and for the individually top-ranked restaurants. That being said, what Table 3 does not capture is whose accessibility is being prioritized, or whose preference. As shown in Table 4, in less than 12% of cases, all members' individually top-ranked locations were actually chosen, with this percentage reducing as clique size increases. Furthermore, irrespective of clique size in around 17% to 20% of cases, no one's top-ranked location was chosen by the clique, suggesting a considerable degree of compromise among members. This underscores the importance of observing the actual decision-making process to gain a better understanding of within-group dynamics.

Table 3. Reasons for choosing restaurant categorized from free answers (multiple criteria allowed)

|  | n | share (%) |
|---|---|---|
| Restaurant quality | 171 | 78.8% |
| Accessibility | 124 | 57.1% |
| Schedule constraints | 25 | 11.5% |
| Voting | 21 | 9.7% |
| Clique dynamics | 8 | 3.7% |
| Inertia | 8 | 3.7% |
| Economic constraints | 1 | 0.5% |
| Others | 12 | 5.5% |

Share was calculated by dividing the n by the number of clique (217)

Table 4. Degree of matching between individually top-ranked locations and clique choice

|  |  | Number of individuals whose top-ranked locations are chosen by the clique | | | | | |
|---|---|---|---|---|---|---|---|
|  |  | 0 | 1 | 2 | 3 | 4 | 5 |
| Clique size | 3 | 18.6% | 37.1% | 33.0% | 11.3% |  |  |
|  | 4 | 14.7% | 29.3% | 28.0% | 18.7% | 9.3% |  |
|  | 5 | 20.0% | 37.8% | 24.4% | 11.1% | 6.7% | 0.0% |

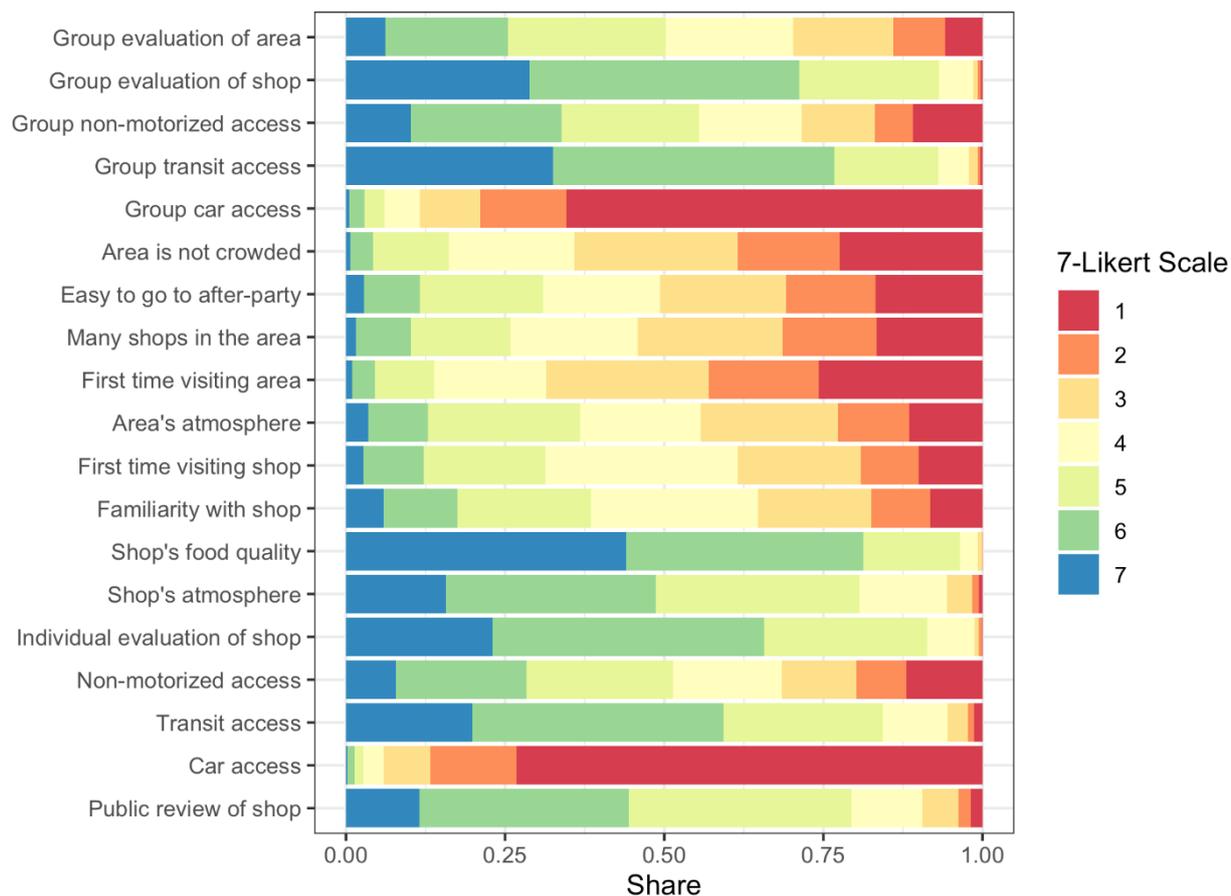

Fig 5. Factors considered important for group-level restaurant choice by individuals (n=816)

### 3.4. Two case studies

To further elucidate the properties of the data collected we will briefly introduce the decision-making process in two particular cases, as summarized in Fig 6 and Fig 7 using information from Surveys 1 to 5 as well as the LINE group discussion text record (a). In particular, the plots of members' schedules (b) and activity places (c) were created using data from the individual preference elicitation survey (Survey 3) and the expected activity diary of the meeting day (Survey 5).

The first clique (Fig 6) is composed of five same-year students (no hierarchy). Two of the members had previous commitments on the suburbs of Tokyo on the day of the activity (1b and 1c). In this particular case several features of the decision-making process can be highlighted (1a). For instance, Mr. A pushed from early in the discussion for his preference, eating French food at Ginza, an upscale district in central Tokyo. Other members, like Mr. C, had a personal preference but showed high degree of agreeableness and willingness to compromise for the group stating: "*My preference is for meat, but if everyone is in for French at Ginza, I don't mind.*" While other alternatives were raised during the discussion such as Japanese BBQ or and oyster bar in Shibuya, Mr. A kept insisting on his preference by posting a link to the shop's online site and menu: "*Let me give you an idea of what French at Ginza will be like.*" It should be noted that most members' individually top-ranked locations were close to their expected origin locations on the day of the activity. Another constraint in the process was that some students were under 20 years old, hence could not drink alcohol, which tilted the choices towards restaurants rather than bars or Japanese izakaya. In the end the group agreed on Mr. A's preference. In this particular case, Mr. A's strong opinion clearly influenced the final decision, given the other member's agreeableness and willingness to compromise. In other words, the weight of Mr. A's opinion was larger than other members. At the same time, we can speculate that had other members had similarly strong opinions, the resulting outcome might have been different. Such information cannot be observed from the outcome alone, but we were able to capture it with the proposed x-GDP method.

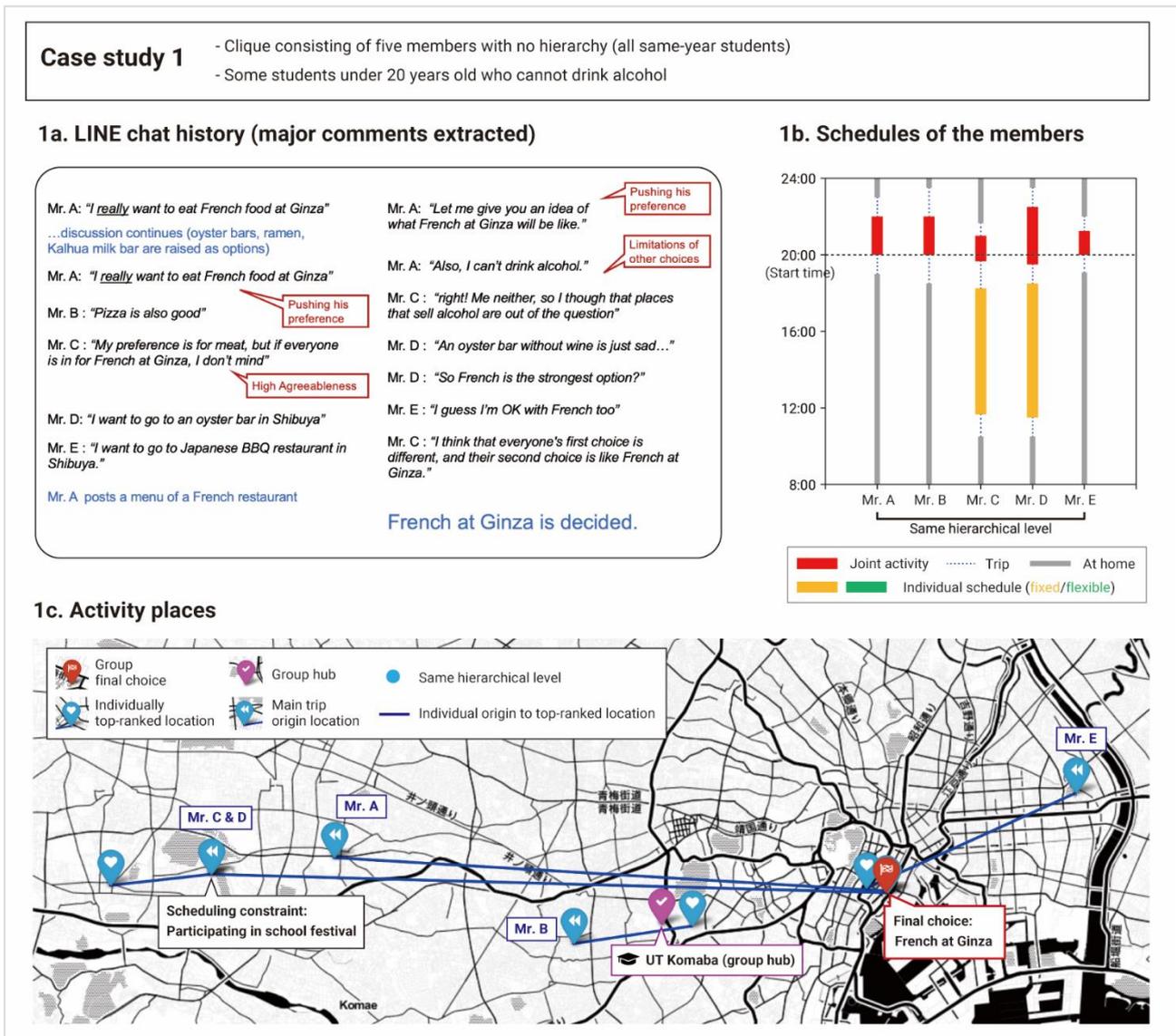

Fig 6. Extract of collected data for a clique (example 1). 1a. LINE chat excerpt. 1b. Schedule of all members on the day of the activity (Survey 5). 1c. OD lines to individually top-ranked location.

The second clique is composed of three futsal club friends, one of them being one year more senior than the other two (two-level hierarchy). First, the joint activity time was set based on two time constraints. First, Mr. A had a part-time job at Shinjuku until 19:00 and second, all members wanted to watch the FIFA World Cup (Qatar 2022) after dinner. Once the time slot was defined, several candidate locations were proposed, but they were all in Shinjuku. A possible reason for this is that

Mr. A had a non-flexible activity schedule during that day. It is also worth noting that Mr. A was the more senior member of the group. The rest of the discussion focused on the restaurant type, such as hotpot, Brazilian BBQ and gibier. In this case, economic constraints were taken into consideration and Brazilian BBQ was selected.

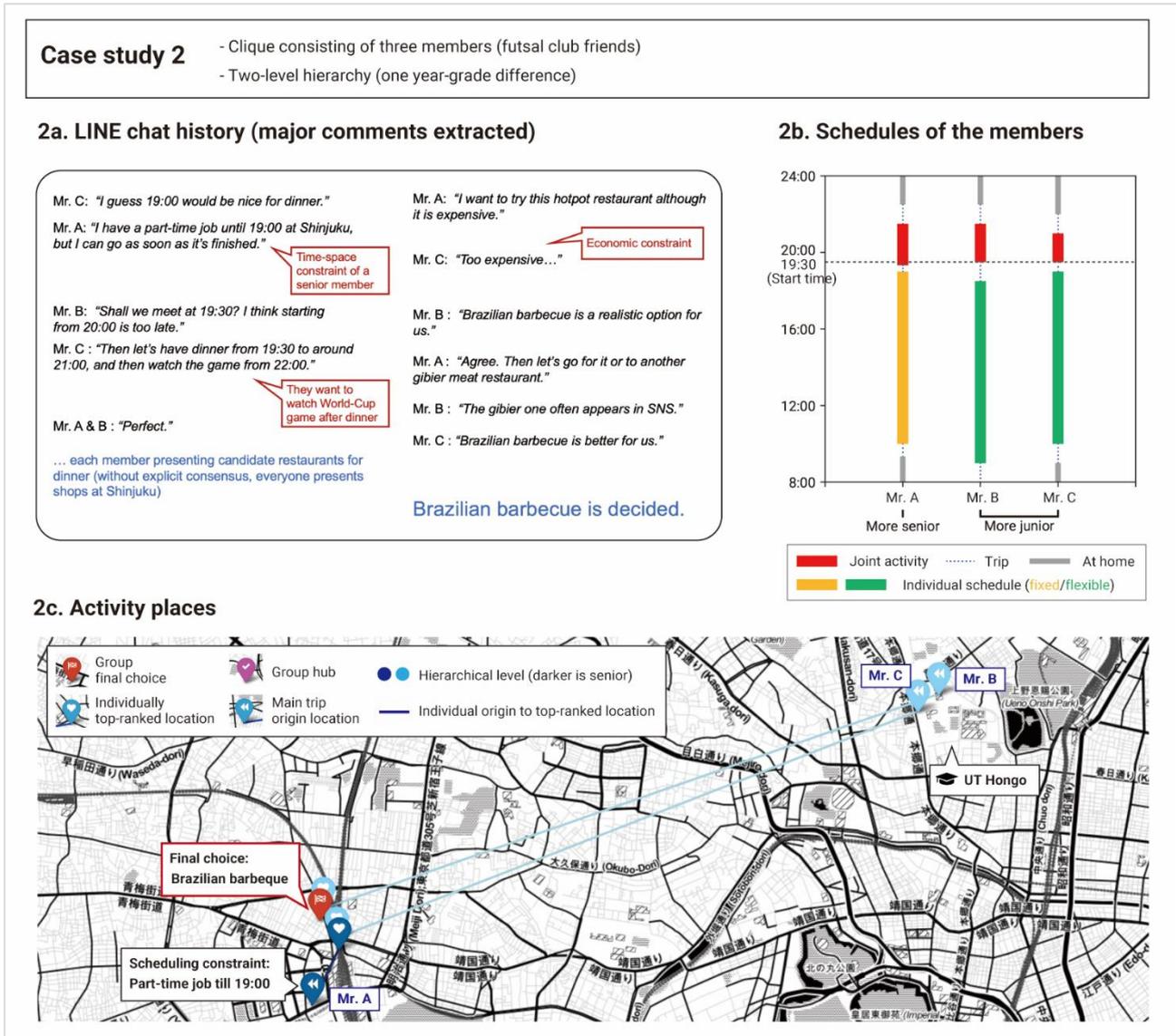

Fig 7. Extract of collected data for a clique (example 2). 2a. LINE chat excerpt. 2b. Schedule of all members on the day of the activity (Survey 5). 2c. OD lines to individually top-ranked location.

4.  Potential avenues of research

In the above sections we have summarized the details of the proposed x-GDP survey method and introduced the results of a survey implementation focusing on joint eating-out activities in the Greater Tokyo Area, giving a detailed overview of the survey components, execution logistics and initial insights on the data. To conclude we want to point out potential avenues of research that can be pursued with this kind of data.

First, we have illustrated with only a few examples, that clique-level decision-making is rather heterogeneous. As such, a first necessary step is a qualitative analysis of the group discussion text records collected to formulate hypothesis regarding decision-making patterns. Such qualitative analysis can be complemented with quantitative methods such as natural language processing and cluster analysis.

Another potential avenue of research is the empirical estimation of joint accessibility and respective parameters. Theoretical joint accessibility methods have been proposed by Neutens *et al.* (2008) however, empirical data is required to estimate model parameters. Joint accessibility estimates can be used to further investigate agglomeration effects in cities, as well as estimate joint activity destination choice models, building on the work of Han *et al.* (2023).

Finally, based on the above, we expect to build a theoretical framework to quantitatively model the joint decision-making process considering clique-level dynamics.

## 5. Acknowledgements

The authors would like to thank Mr. Teruhisa Takizawa, Mr. Koki Okamura and all the Experiment Moderators for their efforts during the experiment execution.
This work was supported by JSPS KAKENHI Grants Number 20H02266.

**Appendix**

Table A1. Individual level characteristics of sample (n=816)

| Variable | n | Share (%) |
|---|---|---|
| **Gender** | | |
|     Male | 630 | 77.21% |
|     Female | 169 | 20.71% |
|     Non-binary / no response | 17 | 2.09% |
| **Grade** | | |
|     First year undergraduate | 77 | 9.44% |
|     Second year undergraduate | 201 | 24.63% |
|     Third year undergraduate | 154 | 18.87% |
|     Fourth year undergraduate | 159 | 19.49% |
|     First year masters | 95 | 11.64% |
|     Second year masters | 80 | 9.80% |
|     Doctoral student | 13 | 1.60% |
|     Other | 37 | 4.53% |
| **Main campus** | | |
|     The University of Tokyo - Hongo | 377 | 46.20% |
|     The University of Tokyo - Kashiwa | 12 | 1.47% |
|     The University of Tokyo - Komaba | 196 | 24.02% |
|     The University of Tokyo - Multiple campuses | 36 | 4.41% |
|     Other universities | 162 | 19.85% |
|     Non-student | 33 | 4.04% |
| **School commuting frequency** | | |
|     6-7 times per week | 118 | 14.46% |
|     4-5 times per week | 414 | 50.74% |
|     2-3 times per week | 184 | 22.55% |
|     1 time per week | 46 | 5.64% |
|     2-3 times per month | 13 | 1.59% |
|     1 time per month | 3 | 0.37% |
|     Less than once per month | 5 | 0.62% |
|     Non-student | 33 | 4.04% |
| **Driver's license** | | |
|     Yes | 561 | 69% |
|     No | 255 | 31% |
| **Car ownership** | | |
|     No | 607 | 74% |
|     Yes, exclusive | 12 | 1% |
|     Yes, shared with other household members | 197 | 24% |
| **Bicycle ownership** | | |
|     No | 259 | 32% |

| | | | | | | |
|---|---|---|---|---|---|---|
| Yes, exclusive | | | 89 | | | 11% |
| Yes, shared with other household members | | | 468 | | | 57% |
| Bike-share usage | | | | | | |
| No | | | 692 | | | 85% |
| Yes, with subscription | | | 8 | | | 1% |
| Yes, without subscription | | | 116 | | | 14% |
| Student/commuter pass | | | | | | |
| Yes | | | 482 | | | 59% |
| No | | | 334 | | | 41% |
| Work style (Includes full time workers and students) | | | | | | |
| Fully remote | | | 131 | | | 16.1% |
| Commuting to fixed location (occasional remote work included) | | | 466 | | | 57.1% |
| Commuting to several locations (occasional remote work included) | | | 19 | | | 2.3% |
| Not working | | | 200 | | | 24.5% |
| Monthly budget for leisure activities (JPY) | | | | | | |
| <10k | | | 52 | | | 6.37% |
| 10k~20k | | | 165 | | | 20.22% |
| 20k~30k | | | 185 | | | 22.67% |
| 30k~40k | | | 173 | | | 21.20% |
| 40k~50k | | | 81 | | | 9.93% |
| 50k~60k | | | 69 | | | 8.46% |
| 60k~70k | | | 43 | | | 5.27% |
| 70k~80k | | | 16 | | | 1.96% |
| 80k~90k | | | 6 | | | 0.74% |
| 90k~100k | | | 10 | | | 1.23% |
| Variable | | n | Median | Mean | S.D. | Min. | Max. |
| Age | | 816 | 21 | 21.7 | 2.71 | 18 | 51 |

Table A2. Public reviews of chosen restaurants and candidates.

| Variable | n | Median | Mean | S.D. | Min. | Max. |
|---|---|---|---|---|---|---|
| Average rating | 1,177 | 3.33 | 3.325 | 0.22 | 3 | 4.33 |
| Number of reviews | 1,182 | 71.5 | 197.71 | 331.32 | 1 | 2,766 |
| Restaurants registered in Tabelog | 99.5% | | | | | |

Of the total 1,188 elicited shops 1,182 were correctly identified in Tabelog. 5 shops did not have any ratings at the time of scraping.

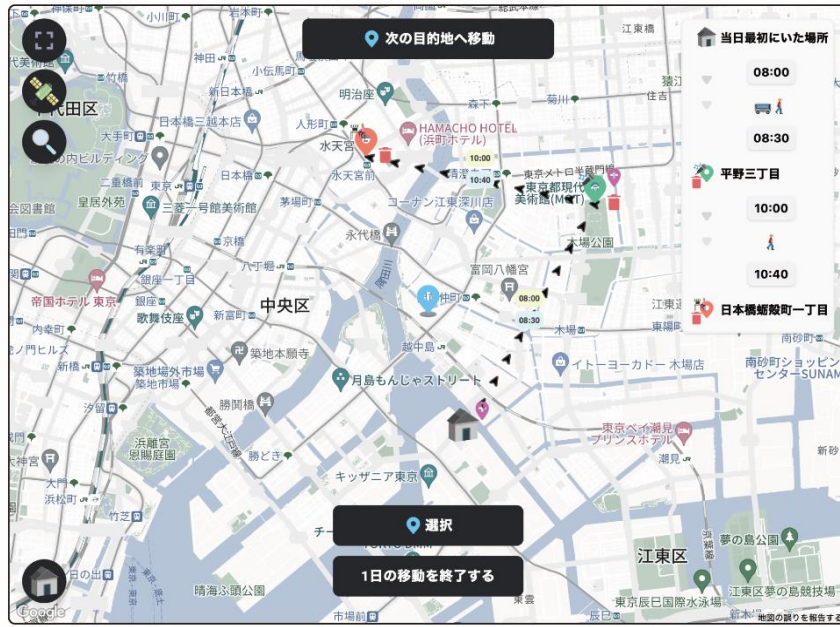
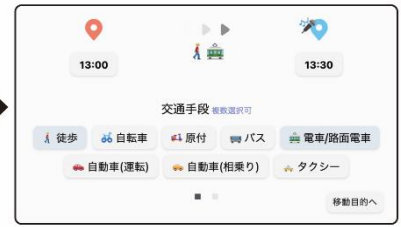
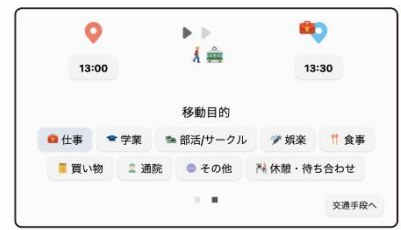

Fig A1. Diary survey and steps of trip registration. Individual first specifies a location (trip destination) on the map, then selects trip arrival and departure times, the mean of transportation, and the trip purpose. The schedule and locations already registered are shown on the map, and the trip starts from the location of the newest record. Edit after registration are also possible.